\documentclass[fleqn,usenatbib]{mnras}

\usepackage[T1]{fontenc}

\DeclareRobustCommand{\VAN}[3]{#2}
\let\VANthebibliography\thebibliography
\def\thebibliography{\DeclareRobustCommand{\VAN}[3]{##3}\VANthebibliography}

\usepackage{graphicx}	
\usepackage{amsmath}	
\usepackage{amssymb}	
\usepackage{tablefootnote}

\usepackage{newtxtext,newtxmath}

\newcommand{\msun}{{\rm ~M}_\odot}	
\newcommand{\kms}{{\rm ~km~s^{-1}}}	
\newcommand{\kpc}{{\rm ~kpc}}	
\newcommand{\sech}{{\rm sech}}	

\title[Local LMC dark matter]{Effects on the local dark matter distribution due to the Large Magellanic Cloud}

\author[Donaldson et al.]{
Katelin Donaldson,$^{1}$\thanks{E-mail: katelinbdonaldson@gmail.com}
Michael S. Petersen,$^{2}$
Jorge Pe{\~n}arrubia$^{1,3}$
\\
$^1$Institute for Astronomy, University of Edinburgh, Royal Observatory, Blackford Hill, Edinburgh EH9 3HJ, UK\\ 
$^2$CNRS and Sorbonne Universite, UMR 7095, Institut d'Astrophysique de Paris, 98 bis Boulevard Arago, F-75014 Paris, France\\
$^3$Centre for Statistics, University of Edinburgh, School of Mathematics, Edinburgh EH9 3FD, UK
}

\date{Accepted XXX. Received YYY; in original form ZZZ}

\pubyear{2021}

\begin{document}
\label{firstpage}
\pagerange{\pageref{firstpage}--\pageref{lastpage}}
\maketitle

\begin{abstract}
We study the local dark matter distribution in two models for the Milky Way (MW)-Large Magellanic Cloud (LMC) interaction. The effect of the LMC on the local dark matter distribution is dependent on the evolution of the MW-LMC system, such that a static model is insufficient to accurately model the dark matter velocity distribution in the solar neighbourhood. An evolved model boosts local LMC dark matter particle velocities by nearly 50\%, to a median value of $\approx750\kms$. MW dark matter particles also experience a velocity boost caused by disc reflex motion owing to the infall of the LMC. We study the implications of LMC particles in the solar neighbourhood for dark matter detection experiments. The directionality of LMC particles is distinguishable from the MW particles, with a difference in the apparent origin centroid location between the MW and LMC particles of $26\pm6 ^\circ$. This unique identifier, along with their high velocities, can be utilised by directional detectors to search for dark matter particles originating in the LMC.
\end{abstract}

\begin{keywords}
Magellanic Clouds -- Galaxy: kinematics and dynamics -- dark matter -- solar neighbourhood -- Galaxy: halo
\end{keywords}



\section{Introduction}

The evasive nature of theorised dark matter particles has fuelled an almost century long search \citep{Zwicky_1937}, with a present goal of verifying the particle nature via direct detection experiments. These experiments conventionally assume that the local dark matter distribution can be characterised by the Standard Halo Model \citep[SHM;][]{Drukier_1986}. The SHM assumes that the velocity distribution is represented everywhere by a smooth Maxwell-Boltzmann distribution by postulating an isothermal dark matter halo profile with $\rho \propto r^{-2}$. However, this assumption has been challenged on a number of occasions, with data sets such as RAVE-TGAS \citep{Herzog_Arbeitman_2018} and SDSS-\emph{Gaia} DR2 \citep{Necib_2019} suggesting anisotropic effects in the local dark matter distribution.

The Large Magellanic Cloud (LMC), a satellite galaxy of the Milky Way (MW), is both a contributor to, and a significant perturber of, the local dark matter distribution. The LMC is currently on its first infall, travelling at $327 \kms$ in a heliocentric frame \citep{Kallivayalil_2013} and has just passed its pericentre, now at a distance of $50$kpc from Earth \citep{Pietrzynski_2019}. The LMC contributes to the local dark matter distribution by populating the high-velocity tail owing to the overlap of its dark matter halo with the solar neighbourhood in standard cosmologically-motivated models \citep{Besla_2019}. Moreover, recent work analysing the dynamics of the MW-LMC system has shown that MW particles in the solar neighbourhood may also have higher velocities than assumed in the SHM \citep{ Petersen..reflexmotion..2020}. These deviations from the SHM provide a unique opportunity for direct detection experiments.

Direct detection experiments probe mass-cross section parameter space for theorised weakly interacting massive particles by reducing background events until a dark matter signal can be detected through interactions with atomic nuclei \citep{Lewin_1996}. Current experiments place exclusion curves on the mass-cross section plane, but as experiment sensitivities increase to the ton-scale \citep[e.g.][]{Aprile_2017}, the next generation of direct detection experiments are approaching an irreducible background signal called the neutrino floor \citep{Boehm_2019}. Neutrinos cause a weak nuclear recoil in the target source, and at low masses, these events can dominate the signal until it is no longer possible to distinguish between the neutrino signal and potential dark matter signal \citep{Grothaus_2014}. The presence of high velocity particles shifts the present exclusion curves to be sensitive to lower masses, further restricting the possible space in which a dark matter signal could be found and increasing the need to reduce the neutrino barrier as it becomes necessary to probe at higher sensitivities and lower masses. However, some direct detection experiments are able to analyse the directionality of the particles detected \citep{Grothaus_2014}, which could aid in reducing the background in both Xenon- \citep{Nygren_2013} and Argon-based \citep{Akimov_2020} detectors.

\vspace{-0.5cm}

\section{Modeling the Milky Way-Large Magellanic Cloud System} \label{Section:Model}

We design idealised cosmologically motivated $N$-body simulations to model the local dark matter distribution. The model consists of a MW comprised of a stellar disc and dark matter halo, and an LMC component comprised of a dark matter halo. Both the MW and LMC halo are modeled using an NFW profile, $\rho(r) = \rho_0\tilde{r}^{-1}(1 + \tilde{r})^{-2}$, where $\rho_0$ is the central density of the dark matter halo, $\tilde{r}=r/R_s$, and $R_s$ is the scale radius of the halo. The MW model has a total mass of $1\times10^{12}\msun$ and roughly matches observational constraints on the mass profile of the MW \citep{Eadie_2019}, with a scale radius $R_{s,{\rm NFW~MW}}=15$kpc.
The MW stellar disc model is an exponential disc with scale length $R_d=3$kpc, a $\sech^2$ vertical profile \citep[$z_0=600$ kpc, in the notation of][]{Petersen_2021_commensurabilities}, and mass $5\times10^{10}\msun$.

We use an LMC virial mass of $M_{\rm vir} = 25 \times 10^{10}\msun$ \citep{Penarrubia_2015,Erkal_2019}. We fix the LMC mass enclosed at 8.7 kpc to match circular velocity observations \citep{vanderMarel_2014}, resulting in $R_{s,{\rm NFW~LMC}}=18$kpc. For both halo profiles, we apply an error function truncation such that the initial halo profile is $\rho_{\rm trunc}(r)=0.5\rho(r)\left(1-{\rm erf}\left[(r-r_{\rm trunc})/w_{\rm trunc}\right]\right)$. The truncation parameters are $r_{\rm trunc}=2R_{\rm vir}$ and $w_{\rm trunc}=0.3R_{\rm vir}$, where $R_{\rm vir,~MW}=300$ kpc, and $R_{\rm vir,LMC}=150$ kpc. 

\begin{figure} 
	\includegraphics[width=\columnwidth]{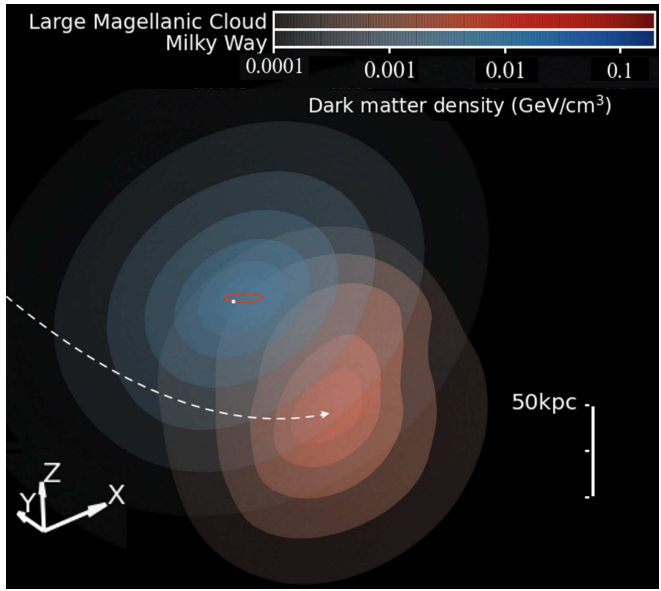}
    \caption{\label{fig:schematic} Present-day dark matter distributions in the MW (blue) and LMC (red). The solar radius is marked with a bright red circle, and the solar neighbourhood with a white marker. The outer isodensity surface ($1\times10^{-4}{\rm GeV}/{\rm cm}^3$) of the LMC dark matter encompasses the solar neighbourhood. Further, the LMC deformation is apparent, elongated along its orbit (white dashed line; the LMC recently passed underneath the MW centre roughly along the $y-z$ plane). The MW deformation is more subtle, and primarily stretches down toward the lower left corner of the figure (the effect is not detectable in the solar neighbourhood). Contours are logarithmically spaced in density, and match between the two components.}
\end{figure}

\begin{table}
 
 \centering
 \begin{tabular}{ccc}
   \hline
     & \multicolumn{2}{c}{Local DM Density (Gev/cm$^3$)}\\
     & Static & Evolved\\
    \hline
    \hline
     MW & 0.30220 & 0.32295 \\
     LMC & 0.00059 & 0.00043 \\
    \hline
    \end{tabular}
 \caption{\label{tab:Densities}The dark matter density at the solar neighbourhood defined as in Section~\ref{Section:Model}.}
\end{table}

We use $N_{\rm MW~disc}=2\times10^6$, $N_{\rm MW~halo}=4\times10^7$, and $N_{\rm LMC}=3\times10^7$ equal-mass particles. The details of the realisation procedure may be found in \citet{Petersen_2021_commensurabilities}. The realisation results in isotropic dark matter halos for the MW and LMC \citep[as in Model 1 of][]{Besla_2019}.

To model the trajectory, we use the LMC centre and mean proper motion for the LMC computed from the measurements in \citet{Kallivayalil_2013}, \(\left(\alpha_{\rm LMC},\delta_{\rm LMC}\right)=\left(78.76^\circ\pm0.52,-69.19^\circ\pm0.25\right)\), \(\left(\mu_{\alpha^\star,{\rm LMC}},\mu_{\delta,{\rm LMC}}\right)=\left(-1.91\pm0.02~{\rm mas/yr},0.229\pm0.047~{\rm mas/yr}\right)\), the distance from \citet{Pietrzynski_2019}, \(d_{\rm LMC} = 49.59\pm0.54~{\rm kpc}\), and the line-of-sight velocity from \citet{vanderMarel_2002}, \(v_{\rm los,~LMC}=262.2\pm3.4~\kms\). The comparison between the simulation and the observed LMC position is done in Cartesian coordinates, and assumes that the peak density of the LMC corresponds to the observed LMC disc centre.
We find the transformation to Cartesian galactocentric coordinates by Monte-Carlo sampling the errors on each of the measurements. We assume $\vec{v}_{\odot\to{\rm LSR}} = (11.1,12.24,7.25)~\kms$ as the Sun’s peculiar velocity with respect to the local standard of rest \citep{Schonrich..2010}, a circular velocity at the solar radius $\vec{v}_{{\rm LSR}\to{\rm GC}}=(0,229,0)\kms$ \citep{Eilers..2019}, the distance to the galactic center as $R_{\rm GC}=8.275~{\rm kpc}$ \citep{Gravity..2021}, and the Sun's height above the galactic midplane as 20.8 pc \citep{Bennett..2019}.

The $N$-body runs are evolved using {\sc exp}, a basis field expansion $N$-body solver \citep{Weinberg..1999,Petersen..EXP..2021} that results in low-noise evolution, with the trade off that the evolution must resemble the allowed degrees of freedom in the system. For the evolution of the MW-LMC system, the basis field expansion is able to parameterise the potential structure at least to the present day. We first run point-mass models of the MW and LMC in reverse from their present-day locations in Cartesian coordinates for $\Delta T=3.5$ Gyr, giving a first trajectory. We then perform a fine-grained initial conditions search around the first trajectory to find a final trajectory that nearly matches the observed LMC phase-space location.  To search the space efficiently, we build a reduced version of the initial conditions by sampling a tenth of the particles from each component. We construct a grid of initial positions and velocities for the LMC that span $[-10\%,0,10\%]$ in $(y,z,v_y,v_z)$. Following results from the initial grid, we refine the initial search space by selecting the closest model and repeating the procedure, this time allowing $(x,v_x)$ to also vary at the $[-3\%,0,3\%]$ level. We then select the closest model from this grid and define the result as our initial position and velocity. Ultimately, we start the LMC at $T=-3.5$ Gyr from $(x,y,z)=(30, 555, -39)~\kpc$ and $(v_x,v_y,v_z)=(0, -96, -30)~\kms$. The MW starts initially at rest. Figure~\ref{fig:schematic} shows a snapshot of the MW and LMC 3-dimensional dark matter distribution at the present day, computed from the basis function expansion. For the static model, we do not allow the MW-LMC system to evolve with time and instead simply place the LMC at the present-day location. While it would be computationally far less expensive to realise a static model, such a model will miss significant dynamical effects caused by the interaction of the MW and LMC. 

We also define a `solar neighbourhood' of particles. Even at the resolution of our current simulations, we are forced to select a large region around the Sun in order to obtain sufficient statistics for analysis. First, based on arguments from the analytic NFW profile, we assume that we will be insensitive to variations on scales smaller than the local value of $|d\rho/\rho|^{-1}$. For our MW model at 8 kpc, this value is 5 kpc. For our LMC model at 50 kpc, this value is 18 kpc. However, we wish to find the smallest radius sphere around the Sun that returns a convergent density value, given our particle sampling. To determine this radius, we define increasing-radius spheres, stepping by $\Delta r=0.5\kpc$ outward from the solar location until we obtain two successive values where the density changes by less than 1\%. We then define this as the systemic error on density measurements in the solar neighbourhood.
Given this study, for the MW, we define the solar neighbourhood as particles with $|r - r_\odot|<2~\kpc$, and for the LMC, we define the solar neighbourhood as particles with $|r - r_\odot|<3~\kpc$. We report the measured densities in Table~\ref{tab:Densities}, highlighting the difference between the static and evolved models: effects induced by the mutual evolution of the MW and LMC.

\vspace{-0.2cm}
\section{Velocity Distributions} \label{Section:VelocityDistribution}

\begin{figure}
    \centering
	\includegraphics[width=\columnwidth]{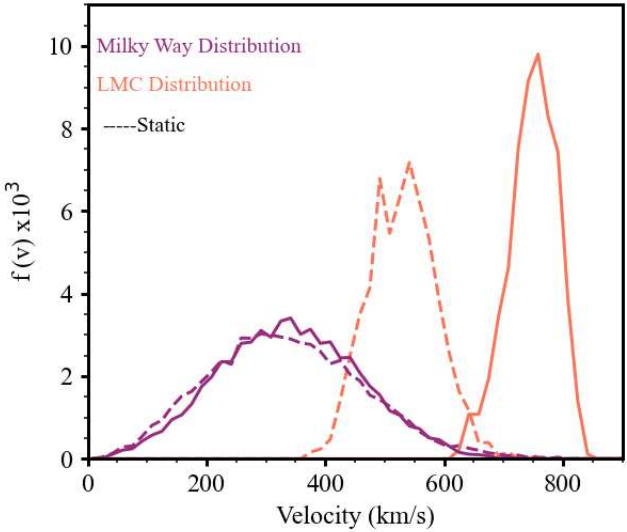}
    \caption{Probability distribution of observed speeds in the solar neighbourhood for the MW and LMC models. The probability distributions are normalised to unity. The dashed lines show results for the static case, while the solid lines show the results in the evolved case, where the MW and LMC have been allowed to affect each other. \label{fig:VelocityDist}}
\end{figure}

Given the simulated particle phase space for the solar neighbourhood, we wish to develop a succinct description of the density and velocity distributions. The density is  directly estimated from the solar neighbourhood (Section~\ref{Section:Model}). The velocity distributions within the solar neighbourhood are more complex, and are the combination of several physical processes. We characterise the velocity of the $i^{\rm th}$ particle with respect to the frame of reference of the Sun (using vector notion where the reference frame between the particle and the Sun is denoted ${}_{\star\to\odot}$) by isolating the contributions from physically motivated dynamical terms, measured with respect to the Milky Way galactic centre (defined as the peak density in the MW; denoted ${}_{\rm GC}$). We write the particle velocity as the sum of vectors,
\begin{equation}
\label{eqn:AllVelocityComponentsMW}
\vec{v}_{i,\star\to\odot}^{\{{\rm MW},{\rm LMC}\}}= \vec{v}_{i,{\rm static}\to {\rm GC}} + \vec{v}_{{\rm GC}\to\odot} + \vec{v}_{i,{\rm perturbed}\to{\rm GC}}.
\end{equation}
The $\vec{v}_{i,{\rm static}\to {\rm GC}}$ term is the intrinsic velocity of each particle, with magnitude governed by the velocity dispersion of the model\footnote{For particles in the LMC, to measure the static models with respect to the MW centre, we need to include both the intrinsic model velocity distribution ($\vec{v}_{i,{\rm static}\to {\rm LMC}}$) as well as the velocity of the LMC centre relative to the MW centre ($\vec{v}_{{\rm LMC}\to{\rm GC}}$): $\vec{v}_{i,{\rm static}\to {\rm GC}}^{\rm LMC} = \vec{v}_{i,{\rm static}\to {\rm LMC}} + \vec{v}_{{\rm LMC}\to{\rm GC}}$. The term $ \vec{v}_{{\rm LMC}\to{\rm GC}}$ is the same for all LMC particles.}. The $\vec{v}_{{\rm GC}\to\odot}$ term is the annual-average velocity of the Sun\footnote{We define the vector $\vec{v}_{\odot\to{\rm GC}}=\vec{v}_{\odot\to{\rm LSR}} + \vec{v}_{{\rm LSR}\to{\rm GC}}$ in Section~\ref{Section:Model}; here we use $\vec{v}_{{\rm GC}\to\odot} = -\vec{v}_{\odot\to{\rm GC}}$. This term is the same between both components, and is the same for every particle.}. 
The $\vec{v}_{i,{\rm perturbed}\to{\rm GC}}$ term covers additional effects, and the characterisation is covered in Sections~\ref{subsubsec:MWdist} (for the MW) and \ref{subsubsec:LMCdist} (for the LMC). 
Each particle velocity vector has an associated speed, $v_i = |\vec{v_i}|$. 
To reduce complexity, we will characterise the speed distribution by each term in equation~(\ref{eqn:AllVelocityComponentsMW}) by reporting the median of the distribution of speeds of the $i$ particles, $\langle v \rangle$. We compute the median separately for the vectors $\vec{v}_{i,\star\to\odot}$, $\vec{v}_{i,{\rm static}\to {\rm GC}}$, and $\vec{v}_{i,{\rm perturbed}\to{\rm GC}}$. These values, computed directly from the simulation for $\langle v_{\star\to\odot}\rangle$ and $\langle v_{{\rm static}\to {\rm GC}}\rangle$, are listed in Table~\ref{tab:PeakVelocities}. After exploring the data, we find that the overall shape of the velocity distributions is set by $\vec{v}_{i,{\rm static}\to{\rm GC}}$, and the additional terms can be modeled as linear-sum contributions\footnote{This formulation is approximate, as it ignores directionality. See the values in Table~\ref{tab:PeakVelocities}. However, most terms are dominated by a single direction, making this approximation roughly true, and instructive for understanding the importance of different terms.}. Therefore, to model the median heliocentric velocity for each component, we only need to have the model velocity dispersion, the annual-average solar motion, and the dynamical ingredients.

\vspace{-0.2cm}
\subsection{Milky Way velocity distribution}
\label{subsubsec:MWdist}

We characterise the dynamical evolution imprint on the MW dark matter particles by dividing the perturbative effects into two terms: 
\begin{equation}
\label{eqn:VelocityComponentsMW}
\vec{v}_{i,{\rm perturbed}\to{\rm GC}}^{\rm MW} = \vec{v}_{i,{\rm secular}\to{\rm GC}} + \vec{v}_{i,{\rm reflex}\to{\rm GC}}.
\end{equation}
The term $\vec{v}_{i,{\rm secular}\to{\rm GC}}$ arises from the deformation of the MW dark matter halo owing to a variety of secular processes \citep[e.g.][]{Petersen_2016} that arise due to the presence of the MW disc. However, the more significant contribution is due to the reflex motion $\vec{v}_{i,{\rm reflex}\to{\rm GC}}$. Reflex motion is characterised by the movement of the MW’s disc relative to the MW barycentre due to the infall of the LMC pulling the MW disc from its equilibrium position at the MW barycentre.
The $\langle v_{{\rm reflex}\to{\rm GC}}\rangle$ value quoted in Table~\ref{tab:PeakVelocities} is measured directly from the motion of the MW disc centre relative to the MW barycentre. The $\langle v_{{\rm secular}\to{\rm GC}}\rangle$ value quoted in Table~\ref{tab:PeakVelocities} is measured from the difference between the initial static model and a control simulation where the LMC is not introduced.

The reflex motion of the MW disc has only recently been measured \citep{Petersen_2021_reflex_motion}, and its inclusion in equation~(\ref{eqn:VelocityComponentsMW}) provides a novel approach to the analysis of the MW dark matter particle velocity distribution. In the heliocentric reference frame, reflex motion causes it to seem as if the outer halo is moving relative to the disc (when in the inertial frame, the disc is moving with respect to the largely at-rest outer halo). Particles with larger apocentres ($r_{\rm apo} \gtrsim 40 \kpc$) experience larger reflex motion velocity effects  \citep{Petersen..reflexmotion..2020}.
Large-apocentre MW particles make up only a small fraction of the particles in the solar neighbourhood, but they have a median velocity higher than the particles in the solar neighbourhood that haven’t experienced reflex motion. Therefore, while the effect of large apocentre particles on the velocity distribution in Figure~\ref{fig:VelocityDist} is relatively small, the particles with the highest heliocentric velocities are more likely to be large-apocentre particles experiencing an apparent velocity boost from reflex motion.

This dynamical scenario is different than the direct acceleration mechanism previously proposed \citep{Besla_2019}. We do not find evidence for any significant direct acceleration of MW particles in our model. The LMC has only recently passed Earth, so local particles have not yet had time to react to the presence of the LMC, which suggests from a timescale argument that direct acceleration is unlikely to be responsible for the high-velocity particles.

\vspace{-0.2cm}
\subsection{Large Magellanic Cloud velocity distribution}
\label{subsubsec:LMCdist}

The motion of the LMC is characterised by the heliocentric velocity of the luminous centre, which is dependent on the initial conditions of the model (see Section \ref{Section:Model}). The heliocentric velocity of the LMC can be measured directly from the simulation and confirmed observationally. 

Figure~\ref{fig:VelocityDist} shows a clear difference in the median velocities of the LMC particles for a static and evolved model, from $\langle v_{\star\to\odot}\rangle=532 \rm \kms$ in the static model, to $\langle v_{\star\to\odot}\rangle=751 \rm \kms$ in the evolved model. It is also clear that in the evolved model, there is only a small overlap with the MW distribution before the high-velocity regime is composed purely of LMC particles. We, therefore, aim to quantify the dynamical effects present in the evolved model that lead to this significant discrepancy. We do this by identifying the unique velocity components contributing to the observed median value. These are described in equation~(\ref{eqn:VelocityComponentsLMC}), where we have specified the vector components of the LMC dark matter particles:
\begin{equation}
\label{eqn:VelocityComponentsLMC}
\vec{v}_{i,{\rm perturbed}\to{\rm GC}}^{\rm LMC} = \vec{v}_{i,{\rm tidal}\to{\rm GC}} + \vec{v}_{i,{\rm reflex}\to{\rm GC}}.
\end{equation}
The term $\vec{v}_{i,{\rm tidal}\to{\rm GC}}$ is the velocity of LMC particles due to the tidal deformation experienced by the LMC and $\vec{v}_{i,{\rm reflex}\to{\rm GC}}$ is the reflex motion\footnote{Using vector notation clarifies a subtle point: the observed infall velocity of the LMC with respect to the galactic barycentre is given by the sum $\vec{v}_{i,{\rm infall}\to{\rm GC}} = \vec{v}_{{\rm LMC}\to{\rm GC}} + \vec{v}_{i,{\rm reflex}\to{\rm GC}}$. That is, when we impose the observed infall velocity vector in the static model, we are implicitly including the reflex motion.}.
The trajectory and distance of the LMC means that all LMC particles will exhibit the maximum reflex motion along their trajectory.

For particles in the solar neighbourhood, $\vec{v}_{{\rm LMC}\to{\rm GC}}$, $\vec{v}_{i,{\rm tidal}\to{\rm GC}}$ and $\vec{v}_{i,{\rm reflex}\to{\rm GC}}$ act to shift the median heliocentric particle velocity approximately as constant offsets owing to their nearly coincident directionality. We neglect $\vec{v}_{i,{\rm secular}\to{\rm LMC}}$ in the LMC case, as we only model the LMC using a spherical distribution. We find that there is a significant velocity boost in the evolved model due to tidal evolution. The tidal acceleration of particles provides a considerable contribution to the observed median velocity of LMC particles, with a median value of $\langle v_{{\rm tidal}\to{\rm GC}}\rangle=225\kms$. The local LMC DM particles are no longer obviously bound to the LMC, but are rather streaming through the solar neighbourhood.

\begin{table}
 \centering
 \begin{tabular}{llcccc}
    \hline
term  && MW$_{\rm static}$   &LMC$_{\rm static}$ & MW$_{\rm evol.}$  &LMC$_{\rm evol.}$\\
    \hline
\multicolumn{2}{l}{$\langle v_{\star\to\odot}\rangle$} & 328 & 532 & 337 & 751 \\
    \hline
\multicolumn{2}{l}{$\langle v_{{\rm static}\to{\rm GC}}\rangle$} & 91  & -  & 91    & - \\
\multicolumn{2}{l}{$\langle v_{{\rm static}\to{\rm LMC}}\rangle$} & -  & 43  & -    & 43 \\
\multicolumn{2}{l}{$\langle v_{{\rm GC}\to\odot}\rangle$} & 242 & 242 & 242   & 242 \\
\multicolumn{2}{l}{$\langle v_{{\rm LMC}\to{\rm GC}}\rangle$} & -  & 323  & -   & 313 \\

\multicolumn{2}{l}{$\langle v_{{\rm perturbed}\to{\rm GC}}\rangle$} & & & & \\
&$\langle v_{{\rm secular}\to{\rm GC}}\rangle$       & 0   & -   & -25   & - \\
&$\langle v_{{\rm reflex}\to{\rm GC}}\rangle$        & 0   & 0   & 51    & 51 \\
&$\langle v_{{\rm tidal}\to{\rm GC}}\rangle$         & -   & 0   & -     & 225 \\
    \hline
    \end{tabular}
 \caption{\label{tab:PeakVelocities} The median speed, in $\kms$, for dark matter particles in the solar neighbourhood with respect to the frame of the Sun (cf. Figure~\ref{fig:VelocityDist}), as well as speed broken down by component following equations~(\ref{eqn:AllVelocityComponentsMW}-\ref{eqn:VelocityComponentsLMC}). }
\end{table}

\vspace{-0.2cm}
\section{Implications for detection} \label{Section:Discussion}

Quantifying the deviation from the SHM in the solar neighbourhood allows for more robust predictions in the search for dark matter, especially with the improving sensitivity of direct detection experiments such as XENON1T \citep{Aprile_2017} and COHERENT \citep{Akimov_2020}. We also discuss the possibility of detecting LMC dark matter with directional detectors and the implications for annual modulation signals.

\vspace{-0.1cm}
\subsection{Direct detection experiments} 
\vspace{-0.2cm}
\begin{figure*}
    \centering
	\includegraphics[width=18cm]{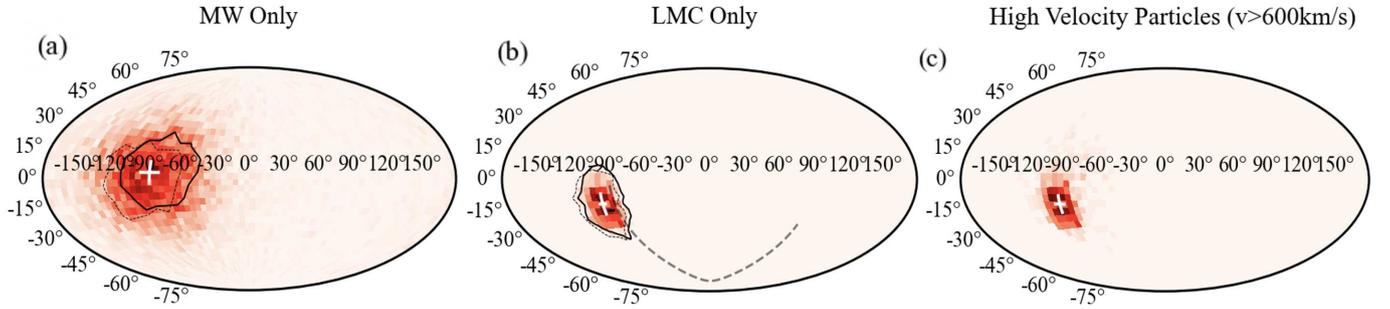}
    \caption{The observed direction of origin for particles plotted on an Aitoff projection, shown only for the evolved model. Darker colours indicate higher numbers of particles. \textbf{(a)} shows only the MW particles \textbf{(b)} shows only the LMC particles \textbf{(c)} shows any particles with a high velocity. The solid curve draws a contour around the densest region in June, and the dotted line shows this in December. The larger dashed curve in  \textbf{(b)}  draws the LMC's infall trajectory (traveling right-to-left).}
    \label{fig:DensityPlot}
\end{figure*}

Previous works have suggested that sources of dark matter substructure might be present in the solar neighbourhood. For example, \citet{Evans_2018} proposed a developed SHM to account for the deviations caused by the Gaia Sausage, and \citet{ O_Hare_2020_substructure} studied the structure resulting from accretion events. The LMC is an obvious significant perturber as well as source of dark matter \citep[as first reported by][]{Besla_2019}, with effects that should not be ignored in the search for dark matter.

In our static model, the MW distribution dominates the signal in the detector, as the LMC density in the static model is negligible compared to the number of MW particles at all values of $v_{\rm min}$. Defining the ratio of densities as a function of velocity, $\eta(v)\equiv \rho_{LMC}(v)/(\rho_{MW}(v)+\rho_{LMC}(v))$), we find that $\eta(v)<0.0015$ for all values of $v$, a value that is not feasibly detectable. However, in the evolved model, the number of MW particles quickly becomes negligible, 99\% of particles are from the LMC at $782\kms$: $\eta(>782\kms)>0.99$. This owes to the appreciably higher median velocity of LMC particles in the evolved model, causing the limited overlap of the  MW and LMC distributions. This creates an opportunity to search for dark matter signatures from LMC particles owing to their high velocities in the heliocentric frame. By creating a $v_{\rm min}$ threshold to create a sample of pure LMC particles, their signature will not be obscured by the population of MW particles. We find that the LMC particles push the distribution to higher velocities, so the exclusion curves will be pushed to lower masses \citep[in agreement with][]{Besla_2019}. 
\vspace{-0.4cm}
\subsection{Directional detectors}
The resolution of direct detection experiments is inherently limited by background events that begin to dominate the signal at some minimum interaction cross section \citep{Gonzalez_Garcia_2018}. Where neutrinos become the dominant background signal, it becomes challenging to distinguish between neutrino signals and dark matter candidates \citep[][]{Boehm_2019}. The next generation of direct detection experiments are aiming to reduce this barrier is by using directional detectors -- detectors sensitive to the apparent origin direction of dark matter particles -- to identify sources of background such as solar neutrinos \citep{Akimov_2020}. 

The distinct trajectory of the LMC provides an exciting opportunity for these detectors. However, it could be challenging due to the fact the LMC's motion is in a similar direction to solar motion ($\vec{v}_{{\rm GC}\to\rm \odot}$). We therefore looked at how sensitive directional detectors would need to be in order to distinguish these as LMC particles. Figure~\ref{fig:DensityPlot} shows the positions of the particles today. The centroid position\footnote{We use the techniques implemented in \href{https://github.com/michael-petersen/unitcentroid}{{\tt unitcentroid}} to compute the centroid of a particle distribution on a unit sphere.} of the observed particle direction origin is given by a white cross, with the size of the arms representing the error bars. The centroid difference between the MW and LMC particles is $26\pm6 ^\circ$. From Figure \ref{fig:DensityPlot}, we conclude that the high velocity particles predominantly come from the LMC, with a significantly reduced scatter in angular position on the sky compared to the MW particles.

Directional detection experiments generally aim to reduce the neutrino barrier by identifying solar neutrinos and removing them from the background counts \citep{Grothaus_2014}. However, there are a number of sources contributing to the neutrino background, not just solar neutrinos, that may complicate this picture. These sources (such as atmospheric, geoneutrinos, and detector neutrinos) are outlined in \citet{O_Hare_2020}. The different sources of neutrinos dominate the background signal of different mass ranges, with solar neutrinos most significant in the low mass range. These alternative neutrino sources contribute to a far more scattered background signal that directional detectors are less equipped to reduce. Therefore, by utilising the unique velocity and directional properties of LMC dark matter particles, we can invert the technique used by directional detectors; instead of searching for particles to reject as background sources, we search directly for possible dark matter signals from the LMC.

\vspace{-0.3cm}
\subsection{Annual modulation}
As well as a change in the count rate in detectors, we also predict variation in the directionality of the particles over the year, induced by the Earth's motion around the Sun. This directional modulation is present in both the MW and LMC particles as shown by the contours in Figure~\ref{fig:DensityPlot}. The average centroid position over the year is shown by a white cross, but the centroid position will change due to annual modulation. Thus we can define the level of sensitivity that directional detectors would need to have to identify this modulation. We find the difference between the direction of the maximum dark matter particles in June and December is $11 \pm 6 ^{\circ}$ for the MW and $4 \pm 4 ^{\circ}$ for the LMC. With modest counts of detections ($\approx10^4$), one can likely detect the modulation in the MW, but the LMC would require significantly more counts in order to constrain the small variations in the maximum direction.

\vspace{-0.4cm}
\section{Conclusions} \label{Section:Conclusions}
We analyse the phase-space distribution of dark matter particles in the solar neighbourhood in both static and evolved models of the MW-LMC system. The main results of this work are as follows:

\begin{enumerate}
    \item Our model predicts LMC dark matter particles crossing the Solar System have a median velocity of $\approx 750 \rm \kms$, with respect to the heliocentric reference frame, which is significantly larger than a model considering only the static velocity of the LMC in this reference frame (Table~\ref{tab:PeakVelocities}, Figure~\ref{fig:VelocityDist}). 
    Current exclusion limits should be revised if the LMC contributes significantly to the local dark matter density at the velocity sensitivity limit of a direct-detection experiment. To get an accurate description of the local phase-space distribution, the evolved model should be considered.
    \item  The increased velocity of LMC particles and decreased density (Table~\ref{tab:Densities}) are primarily due to the LMC tidal evolution (visible in Figure~\ref{fig:schematic}). For MW particles, the effect of reflex motion is more significant than direct acceleration by the LMC in populating the high velocity tail with MW particles. 
    \item The high velocity of LMC particles and directional coherence ($26\pm6 ^\circ$ between the MW and LMC centroids; Figure~\ref{fig:DensityPlot}) in the heliocentric frame provide an opportunity for targeted detection. By searching above a $v_{\rm min}$ threshold to create a sample of pure LMC particles, directional detectors may be able to search for LMC dark matter particles beyond the neutrino floor. 
    \item The centroid change in the annual modulation is $11 \pm 6^{\circ}$ for the MW and $4 \pm 4 ^{\circ}$ for the LMC (Figure~\ref{fig:DensityPlot}). It will therefore be easier to detect annual modulation changes for the MW, compared to the LMC. Owing to the velocity distributions, centroid differences from annual modulation will be easier to detect at lower $v_{\rm min}$ detection thresholds.
\end{enumerate}
The next decade will see an influx of data as direct detection experiments probe for dark matter signatures at higher sensitivities. Our work has shown that, as these experiments progress, it is crucial to account for the presence of the LMC in order to model local phase space accurately. Finally, the study of simulations modelling the MW-LMC system is essential to shed more light on the extent of the LMC's impact, and additional models should be explored.

\vspace{-0.6cm}
\section*{Acknowledgements}
KD thanks the Carnegie Trust for project funding. MSP  acknowledges funding from a French CNRS grant as well as a UK STFC Consolidated Grant. This work used cuillin, the Intitute for Astronomy's computing cluster (http://cuillin.roe.ac.uk), partially funded by the STFC and managed by Eric Tittley. This project made use of \textsc{numpy} \citep{numpy}, \textsc{matplotlib} \citep{matplotlib}, \textsc{ipython} \citep{ipython}, and \textsc{jupyter} \citep{jupyter}.

\vspace{-0.6cm}
\section*{Data Availability}
The data for particles in the solar neighbourhood, including cumulative distribution curves as well as scripts to generate the figures in this paper, is available on Github. \url{https://github.com/katelinbdonaldson/Local-LMC-dark-matter}


\input{thebibliography}


\bsp
\label{lastpage}
\end{document}